\pdfoutput=1
\documentclass[preprintnumbers,amsmath,amssymb,pre]{revtex4}
\usepackage{amsopn}
\usepackage{amsmath}
\usepackage{amssymb}
\usepackage{hyperref}
\usepackage{graphicx}
\usepackage{color}
\usepackage[all]{xy}
\usepackage{subfig}
\usepackage{listings}
\usepackage{subfig}
\usepackage{tikz}
\usetikzlibrary{shapes,snakes}

\newcommand{\ket}[1]{\ensuremath{|#1\rangle}}
\newcommand{\bra}[1]{\ensuremath{\langle#1|}}

\newcommand{\Tr}{\mathrm{Tr}}

\newtheorem{defin}{ Definition}
\newcommand{\boldone}{{\rm 1\hspace*{-0.4ex}%
\rule{0.1ex}{1.52ex}\hspace*{0.2ex}}}

\begin{document}
\fontfamily{serif}

\title{Generating efficient quantum circuits for preparing maximally multipartite entangled states}
\author{Przemys{\l}aw Sadowski}
\affiliation{Institute of Theoretical and Applied Informatics, Polish Academy
of Sciences, Ba{\l}tycka 5, 44-100 Gliwice, Poland}

\begin{abstract}
In this work we provide a method for generating quantum circuits preparing
maximally multipartite entangled states using genetic programming.
The presented
method is faster
that known realisations thanks to the applied
fitness function and several modifications to the genetic programming schema.
Moreover, we enrich the described method by the unique possibility to define an arbitrary structure of a system.
We use the developed method to find new quantum circuits, which are simpler from known
results. We also analyse the efficiency of generating entanglement in
the spin chain system and in the system of complete connections.
\end{abstract}
\maketitle

\section{Introduction}
One of the key resources in the quantum information theory is entanglement
\cite{zyczkowski2006geometry, plenio2007}.
It is an inherent part to many branches of research in the field of quantum information computing.
One of the most perspective applications of quantum computing,
the quantum key distribution, utilizes the entanglement of parts of the
key possessed by a sender and a receiver.
Similarly, the experimental implementation of quantum teleportation
requires sharing of maximally entangled states between the communicating parties.
Entanglement is also necessary in applications such as dense coding and quantum direct
communication protocols.
It is considered that entanglement plays a crucial role in the exponential speedup
of quantum algorithms \cite{jozsa2003}.

The above applications are based on harnessing bipartite maximally
entangled states. This suggests searching for new quantum algorithms
that would make use of maximally multipartite entangled states.
However, searching for maximally entangled states itself is a known problem
\cite{borras2007multiqubit, brown2005searching, tapiador2009simple}.

The main goal of this work is to provide an efficient method of searching for
circuits preparing maximally multipartite entangled states.
The algorithm described in this work is based on genetic programming (GP).
We take advantage of efficient fitness function and introduce the modifications to
the genetic programming engine that result in shorter runtime of the process.
We also introduce the method of representing quantum circuits allowing for
the reduction of the search space.
Additionally we use narrowed sets of universal quantum gates in order to
increase computational efficiency. Apart from improving performance of the
algorithm we also take the structure of a system into consideration.

This paper is organized as follows. In section \ref{sec::gprogramming} we describe
the proposed method, focusing mainly on the circuits representation and the fitness function.
Then, in section \ref{sec::results}, we present the obtained results.

\section{Description of the method}\label{sec::gprogramming}
Genetic programming is the numerical method based on evolutionary mechanisms.
We decided to use such a method because of the two main reasons.
First of all it enables to perform numerical search in complicated, mathematically untraceable space.
On the other hand, we assume that entanglement in a quantum circuits
increases in particular segments quite independently.
Genetic programming enables to exchange segments between different circuits
generating high entanglement by the use of evolution mechanisms.

\subsection{General GP algorithm}
Genetic programming belongs to the family of search heuristics inspired by the mechanism of natural evolution
(in genetic algorithms each element of a search space being candidate for a solution
is encoded as a representative of a population).
Every member of a population has its unique genetic code, which
is its representation in optimization algorithm.
In most of the cases the genetic code is a sequence of values from a finite set $\Sigma$ of
possible values of all the features that characterize a potential solution $x\in\Sigma^n$ in the search space.
Searching for the optimal solution is done by the modification of genetic code due 
according to the rules of
the evolution such as
mutation, selection, crossover and inheritance.

Mutators are the functions that change single elements of a genetic code of a population
member randomly. A basic example of a mutator is a function that randomly changes
values of a representative $x$ at all positions with some non-zero probability

\begin{equation}\label{eq::mutator}
M(x)_i = \left\{
\begin{array}{cc}
x_i,                   & \mathrm{probability}\hspace{0.1cm}p \\
\mathrm{rand}(\Sigma), & \mathrm{probability}\hspace{0.1cm}1-p
\end{array}
\right.\hspace{-0.2cm}.
\end{equation}

Crossovers implement the mechanism of inheritance.
This function divides parental genetic
codes and create a new genetic code.
Commonly two new codes are created at the same time from two parental codes.
An example of such crossover is so called two point cut, where both parental codes ($x_i, y_i$) are cut into three
regions and the middle segments are interchanged

\begin{equation}\label{eq::crossover}
x'_i = \left\{ \begin{array}{cc} x_i & i\le c_1 \lor c_2 \le i \\ y_i & c_1<i<c_2 \end{array}\right.\hspace{-0.15cm},
\hspace{0.5cm}
y'_i = \left\{ \begin{array}{cc} x_i & c_1<i<c_2 \\ y_i & i\le c_1 \lor c_2 \le i \end{array} \right.\hspace{-0.15cm},
\end{equation}
where $c_1< c_2$ are randomly chosen indices.
In every iteration of the algorithm all members of the population are evaluated using fitness function $f:\Sigma^n \to \Re$
which enables ordering elements.
Then, using the selector function, the set of the best members is obtained and used 
to create a new generation of a population using mutation and crossover functions.
There is a number of strategies for defining the selector function --
from completely random choices to the deterministic choice of best representatives.

In this work we use a solution which places somewhere in between.
During every selection we begin with establishing a random set $R_S \subset S $
of elements from population $S$:
\begin{equation}
R_S = rand\{ U\subset S: |U|=n\},  \nonumber
\end{equation}
where $n$ is a fixed number of elements in every set. Then we apply the fitness 
function to choose a maximally well fitted one from this narrowed set (a random element from a set of maximally well fitted ones)

\begin{equation}
 Selector(S) = rand\{ x\in R_S: f(x) = \max_{x\in R_S}  f(x) \},
\end{equation}
where $f(x)$ is the fitness function.

The strategy based on evolution mechanism makes genetic programming especially usable
when parts of genetic code represent the
features of elements of search space and 
can be interchanged between elements independently.
In such case GA is expected to find the features that occur in well fitted representatives
and mix them in order to find the best possible combination.
Pseudo code representing this approach is presented in Listing 1.

\begin{center}

\begin{minipage}{9cm}
\begin{verbatim}
population = RandomPopulation()
for( generationsNumber ){
    newPopulation = []
    for(i = 0; i<population.size()/2; i++){
        mom = Selector(population)
        dad = Selector(population)
        (sister, brother) = CrossOver(mom, dad)
        Mutator(sister)
        Mutator(brother)
        newPopulation.append(sister)
        newPopulation.append(brother)
    }
    population = newPopulation
}
\end{verbatim}
\label{code::gp}
\end{minipage}\\
{Listing 1: }%
Pseudo code representing the algorithm of genetic programming. Functions
{\texttt{ Selector, Mutator} and\texttt{ CrossOver}} work as defined in Section
\ref{sec::gprogramming}.
\end{center}

While the customization of population representation and fitness function unavoidably
relies on the optimization problem,
other parameters of genetic programming such as crossover and mutation methods
are universal.
When treating quantum circuits as the strings of integers representing quantum gates
one can use various, already developed methods \cite{goldberg1989}.

\subsection{Circuits representation - population}
In order to apply the Genetic Algorithm it is
necessary to define a population, a fitness function,the methods of crossing-over,
mutation and selection.
In the case of the optimization of quantum circuits generating entanglement the
population consists of quantum circuits.
The most convenient way to represent a computation in a quantum circuit is a sequence of quantum gates.
Every quantum gate can be approximated using the gates from a set of universal
quantum gates.
In this work we use a set containing the Hadamard gate $H$, the $R(\pi/4)$ (called $\pi/8$)
gate and the controlled-NOT gate $CNOT$ \cite{nielsen2002quantum}

\begin{equation} \label{eq::gates}
H=\frac{1}{\sqrt{2}}\left( \begin{array}{cc}
1 & 1 \\
1 & -1
\end{array} \right),
CNOT=\left( \begin{array}{cccc}
1 & 0 & 0 & 0 \\
0 & 1 & 0 & 0 \\
0 & 0 & 0 & 1 \\
0 & 0 & 1 & 0
\end{array} \right),
R(\pi/4)=\left( \begin{array}{cc}
1 & 0 \\
0 & e^{i\pi/4}
\end{array} \right).
\end{equation}

However, initial experiments suggest that the $R(\pi/4)$ gate is not necessary
for generating maximally entangled states in small systems (i.e. with less than 7 qubits).
Thus, in order to simplify the resultant state,
we consider a set containing only $H$ and $CNOT$ gates when the obtained entanglement is maximal.

One should note that using a general representation of circuits, we treat circuits that are equivalent with respect to
permutations of qubits as completely different.
In out scheme to decrease the number of equivalent circuits in the search space we reduce the
number of available gates on a few first
positions to the ones operating on first $2 p$ qubits, where $p$ is the number of a gate.
Such simplification does not influence the space of feasible solutions, because
it is not possible to act on more than $2 p$ qubits using $p$ gates which all
act on maximum $2$ qubits.

\subsection{Multipartite entanglement potential - fitness function}
As our goal is to find a circuit which prepares maximally multipartite entangled
state, our fitness function must determine how entangled the resultant
state is.
In this case computing fitness score of a quantum circuit requires obtaining the output
state and then the estimation of its entanglement.
One should note that we do not necessarily need an entanglement measure
to perform a successful numerical optimization.
We only need a function for some kind of estimation of multipartite entanglement
that reaches its maximum only for maximally entangled states, which can be characterized
by a number of criteria.
\begin{defin}
State $\rho\in D(\mathcal{X}\otimes \mathcal{Y})$, $\dim(\mathcal{X})\ge\dim(\mathcal{Y})=n$ is maximally entangled if and only if the reduced state is maximally mixed: $\Tr_\mathcal{X}(\rho)=\boldone/n$ i.e.
$S(\Tr_\mathcal{X}(\rho))=n$.
\end{defin}

As we do not have a good measure of multipartite entanglement that we can use in our computation,
we try to find a function that estimates the entanglement quantitatively.
In this work we rely on the definition of maximally multipartite entangled states
introduced by Facchi et al. \cite{facchi2008maximally}.
\begin{defin}
State $\ket{\psi}\in\mathbb{C}^{2^n}$ is called maximally multipartite entangled state if it is maximally entangled according to every bipartition.
\end{defin}

That definition induces a construction for multipartite entanglement potential, 
which is based on the algebraic sum of the purity of all possible bipartitions of a system.
Such an approach guarantees that the maximally entangled states are classified correctly.
However, the purity function does not distinguish some states with different values of
entanglement measures such as negativity or von Neumann entropy.
For this reason in this work we use analogous algebraic sum of bipartite entanglement measure defined as

\begin{equation}
E_{VN}^{(n)}(\ket{\psi}\bra{\psi}) = \sum_{(\mathcal{X},\mathcal{Y})}E_{VN}({\Tr}_{\mathcal{X}}(\ket{\psi}\bra{\psi})) ,
\end{equation}
where $(\mathcal{X}, \mathcal{Y})$ is a bipartition of a system, and $\dim(\mathcal{X})\ge\dim(\mathcal{Y})$.
For pure states it holds that $\Tr_Y(\ket{\psi}\bra{\psi})=AA^\dagger$,
where $A$ is the coefficients matrix of a state $\ket{\psi}$ (i.e. $\ket{\psi}=\mathrm{vec}(A)$).
Thus we have:
\begin{equation}
E_{VN}(\ket{\psi}\bra{\psi}) = E_d(\ket{\psi}\bra{\psi}) = S(\Tr_Y(\ket{\psi}\bra{\psi})) = S(AA^\dagger) =-\sum\lambda_i\log_2\lambda_i,
\end{equation}
where S is the von Neumann entropy,  and $\lambda_i$ are  of the matrix $A$.

\begin{figure}
\includegraphics{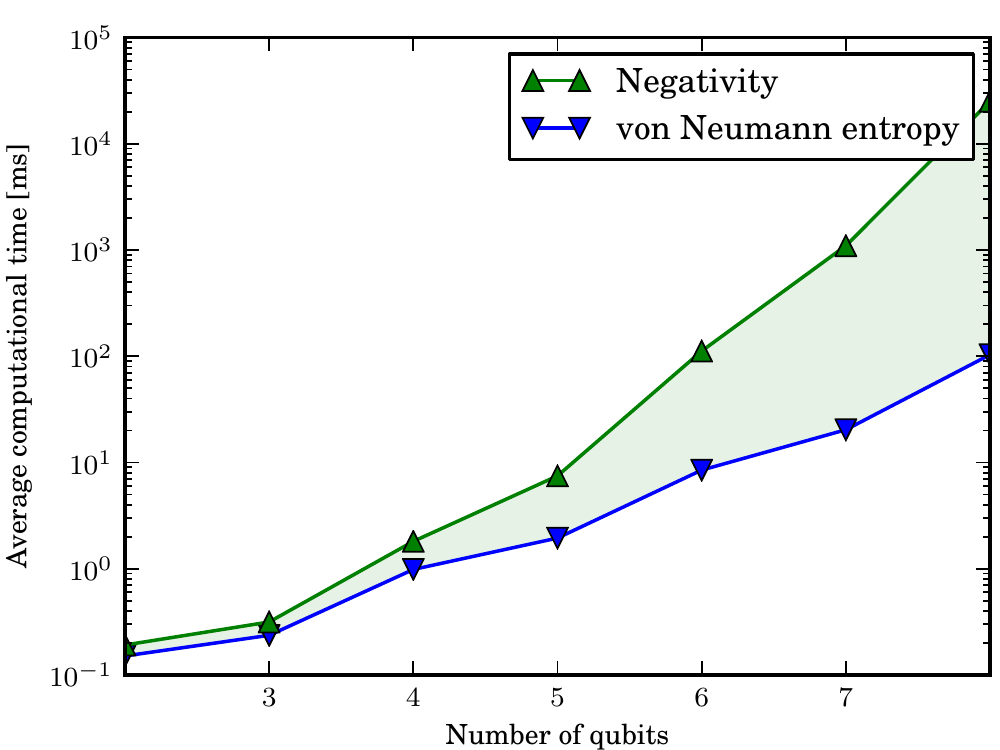}
\caption{Average computational time (log scale) required to calculate negativity
and
von Neumann entropy. For each dimension
$d=1,\ldots,8$ the results were obtained for a sample of $10^4$ random states.
}
\label{fig::measures}
\end{figure}

Although we considered a number of functions,
only proper bipartite measures of entanglement have met our restrictions.
In order to select one for further computation we performed
computational speed
comparison (see Figure \ref{fig::measures}).
Finally
we decided to use algebraic sum of a measure using von Neumann Entropy over all possible
bipartitions of a system.

%\newpage
\section{Generation of multipartite entanglement}\label{sec::results}
As an example of an application of the described method we find the minimal circuits preparing the
maximally multipartite entangled states.
We perform search both for spin chain and complete connection systems.
Finally we compare the obtained results.
The method developed in this paper is probabilistic.
This makes the optimization of additional parameters a tedious task.
In order to obtain a minimal circuit we execute an algorithm with increasing number of gates.
We stop when a maximally entangled state is prepared in a circuit of a fixed size.
When the hypothetical maximum of the fitness function is unreachable,
we search for the maximal value of the entanglement potential and
then optimize the size of a circuit. 
In this work we restrict ourselves to 8 qubits.
This is caused by the long runtimes of the algorithm.

The optimization is preformed simultaneously for the spin chain and the system of complete connections.
By a completely connected system we mean a system in which we are able to act on every pair
of qubits.
It is equivalent to defining the set of all available CNOT gates as
$\{CNOT(i; j) : i \ne j \}$, where $CNOT(i,j)$ is a controlled NOT gate acting on $i$-th and $j$-th qubit (\ref{eq::gates}).
The sets of Hadamard gates and phase gates are independent of changes of
possible qubit connections as this gate acts on one qubit only.
By the spin chain system we
mean a system where we are able to act on pairs of the nearest neighbors in a chain.
In such a case the set of available CNOT gates is $\{CNOT(i; j) : |i-j|=1\}$.
In each complete system we additionally analyze the minimum topology of necessary connections
by investigating the connections graph where the vertices represent qubits of a system, and the edges represent
connections, that is CNOT gates of a circuit connecting two qubits.

\vspace{0.6cm}
\begin{table}

\begin{center}

\begin{tikzpicture}[x=.75cm,y=.5cm]
\draw (4.5,0) grid [step=1.5] (13.5,4.5);
\draw (0,4.5) -- (4.5,3.0);
\draw (0,0.0) rectangle (4.5,1.5);
\draw (0,1.5) rectangle (4.5,3.0);
\draw (0,3.0) rectangle (4.5,4.5);

\node at (5.25,3.1) [above ,inner sep=1pt] {3};
\node at (6.75,3.1) [above ,inner sep=1pt] {4};
\node at (8.25,3.1) [above ,inner sep=1pt] {5};
\node at (9.75,3.1) [above ,inner sep=1pt] {6};
\node at (11.25,3.1) [above ,inner sep=1pt] {7};
\node at (12.75,3.1) [above ,inner sep=1pt] {8};

\node at ( 5.25 ,1.6) [above ,inner sep=1pt] {3};
\node at ( 6.75 ,1.6) [above ,inner sep=1pt] {5};
\node at ( 8.25 ,1.6) [above ,inner sep=1pt] {8};
\node at ( 9.75 ,1.6) [above ,inner sep=1pt] {12};
\node at ( 11.25 ,1.6) [above ,inner sep=1pt] {18};
\node at ( 12.75 ,1.6) [above ,inner sep=1pt] {30};

\node at ( 5.250 ,0.1) [above ,inner sep=1pt] {3};
\node at ( 6.75 ,0.1) [above ,inner sep=1pt] {5};
\node at ( 8.250 ,0.1) [above ,inner sep=1pt] {10};
\node at ( 9.75 ,0.1) [above ,inner sep=1pt] {17};
\node at ( 11.250 ,0.1) [above ,inner sep=1pt] {30};
\node at ( 12.75 ,0.1) [above ,inner sep=1pt] {40};
\node at (0.5,1.5) [above right,inner sep=1pt] {Spin chain};
\node at (0.5,0.0) [above right,inner sep=1pt] {Complete};
\node at (2.8,3.5) [above right,inner sep=1pt] {qubits};
\node at (0.5,3.0) [above right,inner sep=1pt] {System};
\end{tikzpicture}

\end{center}
\caption{The minimal number of all gates necessary to generate maximum multipartite entanglement in spin chain and system of complete connections\label{tab::gates}}

\end{table}

\begin{table} 

\begin{center}

\begin{tikzpicture}[x=.75cm,y=.5cm]
\draw (4.5,0) grid [step=1.5] (13.5,4.5);
\draw (0,4.5) -- (4.5,3.0);
\draw (0,0.0) rectangle (4.5,1.5);
\draw (0,1.5) rectangle (4.5,3.0);
\draw (0,3.0) rectangle (4.5,4.5);

\node at (5.25,3.1) [above ,inner sep=1pt] {3};
\node at (6.75,3.1) [above ,inner sep=1pt] {4};
\node at (8.25,3.1) [above ,inner sep=1pt] {5};
\node at (9.75,3.1) [above ,inner sep=1pt] {6};
\node at (11.25,3.1) [above ,inner sep=1pt] {7};
\node at (12.75,3.1) [above ,inner sep=1pt] {8};

\node at ( 5.25 ,1.6) [above ,inner sep=1pt] {2};
\node at ( 6.75 ,1.6) [above ,inner sep=1pt] {3};
\node at ( 8.25 ,1.6) [above ,inner sep=1pt] {5};
\node at ( 9.75 ,1.6) [above ,inner sep=1pt] {8};
\node at ( 11.25 ,1.6) [above ,inner sep=1pt] {11};
\node at ( 12.75 ,1.6) [above ,inner sep=1pt] {16};

\node at ( 5.250 ,0.1) [above ,inner sep=1pt] {2};
\node at ( 6.75 ,0.1) [above ,inner sep=1pt] {3};
\node at ( 8.250 ,0.1) [above ,inner sep=1pt] {7};
\node at ( 9.75 ,0.1) [above ,inner sep=1pt] {11};
\node at ( 11.250 ,0.1) [above ,inner sep=1pt] {18};
\node at ( 12.75 ,0.1) [above ,inner sep=1pt] {23};

\node at (0.5,1.5) [above right,inner sep=1pt] {Spin chain};
\node at (0.5,0.0) [above right,inner sep=1pt] {Complete};
\node at (2.8,3.5) [above right,inner sep=1pt] {qubits};
\node at (0.5,3.0) [above right,inner sep=1pt] {System};
\end{tikzpicture}
\end{center}

\caption{The minimal number of CNOT gates necessary to generate maximum multipartite entanglement in spin chain and system of complete connections\label{tab::cnots}}

\end{table}

\subsection{Resultant states and circuits}
The upper bound of the potential of entanglement can be given explicitly
if the number of all possible bipartitions is known.
However, it is not always possible to find a pure state with the
maximum value of the entanglement potential.

We have managed to reach the hypothetical maximum in the case of 3, 5 and 6 qubits.
These are the most important cases, because there are no concerns about
the correctness of their classification by the fitness function.
3- and 5-qubit states obtained by us are analogous to the states
presented in \cite{fan2010constructing}.
In the 6 qubit case the best known state was presented in work \cite{fan2010constructing}.
However, that state consists of 32 non-zero coefficients.
We have managed to find a state with 16 non-zero coefficients

\begin{equation}
\begin{split}
\ket{\psi_6}=&\frac{1}{4}
((\ket{0000}-\ket{1111})\ket{\psi^+} + (\ket{0011}+\ket{1100})\ket{\psi^-}+\\&
(\ket{0101}+\ket{1010})\ket{\phi^+}+(\ket{0110}-\ket{1001})\ket{\phi^-}),
\end{split}
\end{equation}
where $\ket{\psi^\pm}=\ket{00}\pm\ket{11}$ and $\ket{\phi^\pm}=\ket{01} \pm \ket{10}$.

In the case of 4 and 7 qubits it is not known whether there is any state with hypothetical maximum of
the potential of entanglement and in 8-qubit case it is definitely not possible.
Thus we can only search for circuits that generate the amount of entanglement that we consider
maximally possible.
The values obtained in our research are gathered in Table \ref{tab::valuesEnt}.

\begin{table}
\begin{center}

\begin{tikzpicture}[x=.75cm,y=.5cm]
\draw (4.5,0) grid [step=1.5] (13.5,4.5);
\draw (0,4.5) -- (4.5,3.0);
\draw (0,0.0) rectangle (4.5,1.5);
\draw (0,1.5) rectangle (4.5,3.0);
\draw (0,3.0) rectangle (4.5,4.5);

\node at (5.25,3.1) [above ,inner sep=1pt] {3};
\node at (6.75,3.1) [above ,inner sep=1pt] {4};
\node at (8.25,3.1) [above ,inner sep=1pt] {5};
\node at (9.75,3.1) [above ,inner sep=1pt] {6};
\node at (11.25,3.1) [above ,inner sep=1pt] {7};
\node at (12.75,3.1) [above ,inner sep=1pt] {8};

\node at ( 5.25 ,1.6) [above ,inner sep=1pt] {3};
\node at ( 6.75 ,1.6) [above ,inner sep=1pt] {10};
\node at ( 8.25 ,1.6) [above ,inner sep=1pt] {25};
\node at ( 9.75 ,1.6) [above ,inner sep=1pt] {66};
\node at ( 11.25 ,1.6) [above ,inner sep=1pt] {154};
\node at ( 12.75 ,1.6) [above ,inner sep=1pt] {372};

\node at ( 5.250 ,0.1) [above ,inner sep=1pt] {3};
\node at ( 6.75 ,0.1) [above ,inner sep=1pt] {9.0};
\node at ( 8.250 ,0.1) [above ,inner sep=1pt] {25};
\node at ( 9.75 ,0.1) [above ,inner sep=1pt] {66};
\node at ( 11.250 ,0.1) [above ,inner sep=1pt] {151.6};
\node at ( 12.75 ,0.1) [above ,inner sep=1pt] {363.1};
\node at (0.5,1.5) [above right,inner sep=1pt] {Upper bound};
\node at (0.5,0.1) [above right,inner sep=1pt] {Obtained value};
\node at (2.8,3.5) [above right,inner sep=1pt] {qubits};
\node at (0.5,3.1) [above right,inner sep=1pt] {Value};
\end{tikzpicture}

\end{center}

\caption{The maximum value of the entanglement potential obtained for $d=1,\ldots,8$ qubits and hypothetical upper bound\label{tab::valuesEnt}}
\end{table}

\definecolor{mycolor}{RGB}{0,128,0}

\subsection{Systems comparison}
The main feature of a quantum circuit that gives us some information about the
generated state is its size.
In our work we put the main interest
in the minimal number of quantum gates needed to obtain a circuit which manages to
generate a maximally entangled state.
The number of basic gates needed to generate maximum entanglement provide
insight into the difficulty of this process.

\begin{figure}
\includegraphics{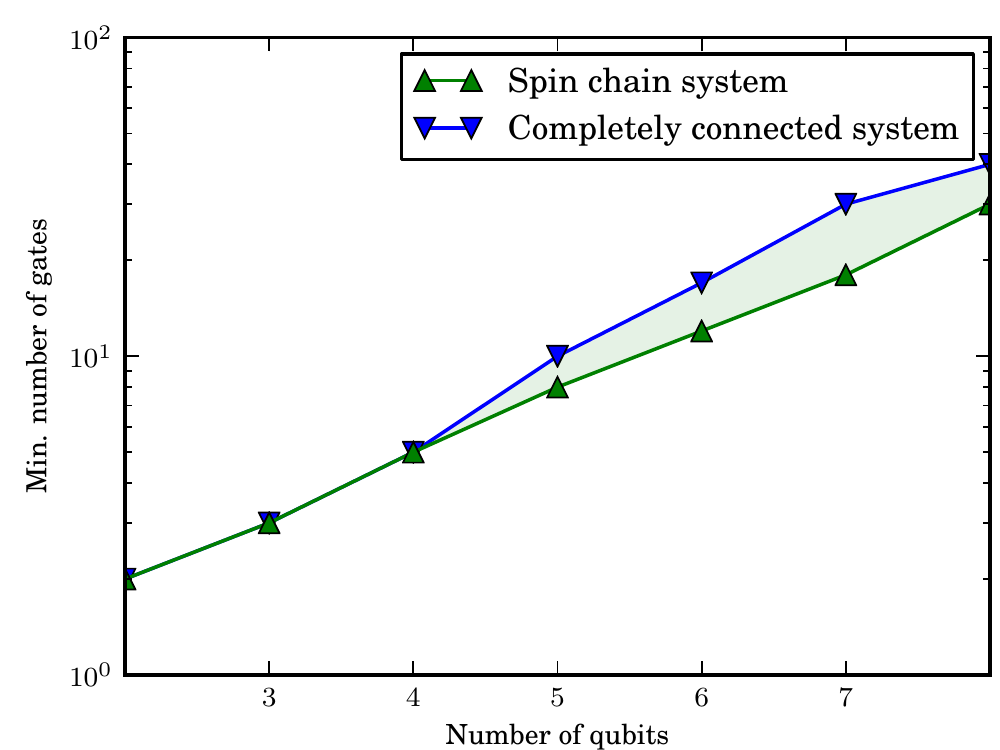}
\caption{The minimum number of gates (log scale) necessary for generating maximally (highly)
entangled states.\label{fig::comparison}}
\end{figure}

All obtained data are gathered in Tables \ref{tab::gates} and \ref{tab::cnots}. 
The computation performed by us shows that the size of the circuit grows
exponentially with the size of a system (see Figure \ref{fig::comparison}).
If we treat the gates used in our work as the elementary operations, we find out that
the complexity of all quantum algorithms that harness
the generation of multipartite entanglement is exponential.
It means that all hypothetical quantum algorithms which are supposed to
bring essential computational speed-up by utilising multipartite entanglement
may be practically inefficient.

Additionally, considering both complete systems and spin chains,
the obtained results show that the simulation of dynamics of a complete system
using spin chain system brings an exponential growth of the number of quantum gates needed.
That may suggest that quantum systems of informatics based on spin chains
may be exponentially less efficient than the systems where the interactions
between all pairs of qubits are possible.

\subsection{Optimal structures}
\begin{figure}[h]
	\centering
	\subfloat[5 qubits]{%
		$$
		\xymatrix{%
		{} & {} & q1 \ar[dr] \ar[dl] {} & {} & {} \\
		q4 \ar[r] & q2 \ar[l] \ar[rr] \ar[ur] & {}  & q3 \ar[ll] \ar[lu] \ar[r] & q5 \ar[l] \\
		}%
		$$\label{subfig::5qconns}
	}\\
	\subfloat[6 qubits]{
		$$
		\xymatrix{
		{} & {} & q1 \ar[d] & {} & {} \\
		{} & {} & q2 \ar[dr] \ar[dl] \ar[u] & {} & {} \\
		{} & q3 \ar[ld] \ar[rr] \ar[ur] & {}  & q4 \ar[rd] \ar[ll] \ar[lu] & {} \\
		q5 \ar[ur] & {} & {} & {} & q6 \ar[ul] 
		}
		$$
	}\\
	\subfloat[7 qubits]{
		$$
		\xymatrix{
		{} & {} & {} & q1 \ar[d] \ar[dddlll] \ar[dddrrr] & {} & {}  & {} \\
		{} & {} & {} & q2 \ar[dr] \ar[dl] \ar[u] & {} & {}  & {} \\
		{} & {} & q3 \ar[lld] \ar[rr] \ar[ur] & {}  & q4 \ar[rrd] \ar[ll] \ar[lu] & {}  & {} \\
		q5 \ar[urr] \ar[uuurrr] & {} & {} & q7 \ar[rrr] & {}  & {} & q6 \ar[ull] \ar[uuulll] \ar[lll]
		}
		$$
	}
	\caption{Connection graphs of the optimal (a) 5-qubit  (b) 6-qubit and (c) 7-qubit quantum circuit preparing maximally entangled state.\label{fig::connectionsAll}}
\end{figure}
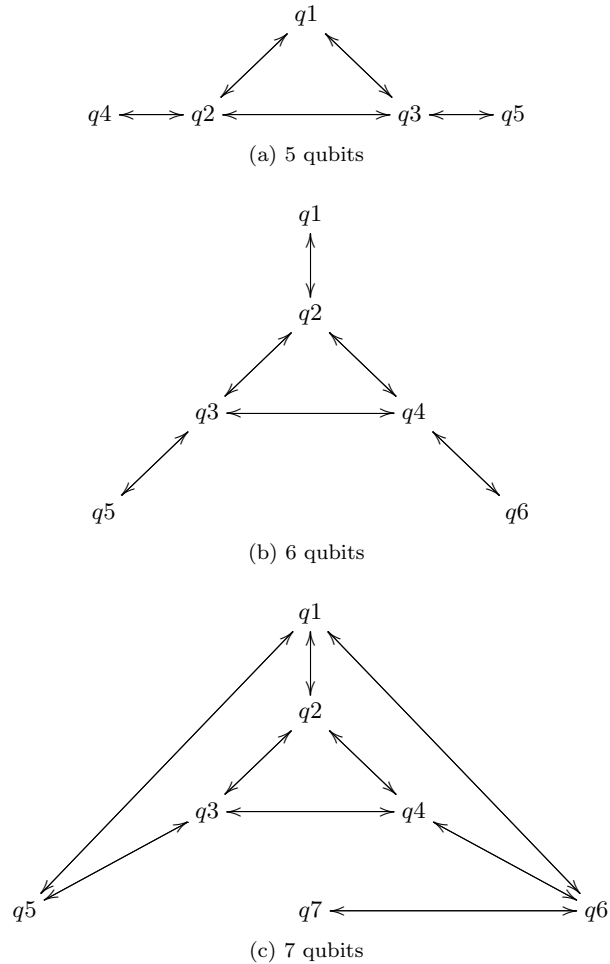

In our work we additionally analyse the structure of quantum systems for the presented circuits.
While spin chain systems seem to be inefficient in the context of generating MME we find it interesting
to find a structure that guarantees maximum efficiency without complete connections.
The results shown in Figure \ref{fig::connectionsAll}
represent the structures of systems allowing the
generation of maximum entanglement in an optimal time.
These examples show that it is not necessary to allow the interaction between all qubits to
obtain maximum efficiency.
We can get an essential reduction of qubit connections without increasing
the number of gates necessary to generate maximum entanglement.
The evolution of the structure seems to be regular.

\section{Concluding remarks}\label{sec::conclusions}
In this paper we have presented the method of applying the genetic algorithm to generate
quantum circuits preparing maximally entangled states.
The algorithm provided in this work can be applied to optimize the size of a circuit
preparing maximally multipartite entangled states.
We have provided an analysis of minimum number of quantum gates necessary for
generating MMES.
The size of a circuit is interesting both 
in the context of minimal complexity of a circuit generating the maximum entanglement and
in obtaining maximally entangled states itself,
because
the complexity of the algebraic representation of a resulting state increases
with the number of gates in a circuit.
The circuit we have obtained for a 6-qubit system that contains $12$ quantum gates and generates a maximally
entangled state with 16 non-zero coefficients is an example.
The best result known so far included a $13$-gate circuit generating a state with $32$ coefficients  \cite{fan2010constructing}.

Moreover, the described method allows taking into account the topology of connections
between the particles in the system.
Developed method enables one to
ensure that the resultant circuit will be possible to implement in a system
with arbitrarily defined connections between particles.
This gives us the possibility of analysis the process of
generation of the multipatrtite entanglement.
We exploit this possibility and provide the comparison of efficiency of generating
maximally multipartite entangled states in the spin chain system and in the system of
complete connections.
We have shown that in the case of generation the multipartite entanglement
a spin chain is essentially less efficient than the system of
complete connections.
Moreover, we have presented some topologies with relatively sparse connections
between qubits that enable the generation of maximum entanglement with the same
efficiency as the systems of complete connections.

\section{Acknowledgements}
This work was supported by the National Science Centre under the research
project DEC-2011/03/D/ST6/00413. The author would like to acknowledge
interesting and motivating discussions with J. Miszczak.

\bibliography{genetic}
\bibliographystyle{apsrev}
\end{document}